\documentclass[aps]{revtex4}

\usepackage[dvips]{color}
\usepackage{times}
\usepackage{amssymb}

\newcommand{\BCH}{Baker-Campbell-Hausdorff }
\newcommand{\e}{\textnormal{e}}
\newcommand{\NN}{\mathbb{N}}

\begin{document}

\title{A note on the Zassenhaus product formula}

\author{Daniel Scholz}
\affiliation{Institut f{\"u}r Theoretische Physik, Universit{\"a}t
G{\"o}ttingen, Tammannstra{\ss}e 1, D-37077 G{\"o}ttingen, Germany.}
\email{mehr-davon@gmx.de}

\author{Michael Weyrauch}
\affiliation{Physikalisch-Technische Bundesanstalt, Bundesallee 100,
D-38116 Braunschweig, Germany.}
\email{michael.weyrauch@ptb.de}

\date{\today}

\begin{abstract}
We provide a simple method for the calculation of the terms $c_n$ in
the Zassenhaus product
$\e^{a+b}=\e^a\cdot\e^b\cdot\prod_{n=2}^{\infty}\e^{c_n}$ for
non-commuting $a$ and $b$. This method has been implemented in a
computer program. Furthermore, we formulate a conjecture on how to
translate these results into nested commutators. This conjecture was
checked up to order $n=17$ using a computer.

\end{abstract}

%\pacs{02.20.Sv, 02.30.Mv}

\maketitle

\section{Introduction}

The product of the exponentials of two non-commuting variables $x$
and $y$ may be expressed in terms of the \BCH (BCH) series
\begin{equation}\label{BCH}
  \e^x \e^y~=~\e^{x+y+\sum_{n=2}^\infty z_n}.
\end{equation}

%\limits_{n=2}^{\infty}
The terms $z_n$ of the sum may be written as linear combinations of
words $W$ of length $n$ consisting of letters $x$ and $y$,
\begin{equation}\label{BCH-terms}
  z_n~=\sum_{W(s_1,\ldots,s_n)} z(W)~ W(s_1,\ldots,s_n),
\end{equation}
where each word $W(s_1,\ldots,s_n)$ is a product of $n$ factors
$s_i=x$ or $s_i=y$. The sum in Eq.~(\ref{BCH-terms}) is over all
possible different words, i.e. the sum over $W$ has in principle
$2^n$ terms. Some of these terms vanish, because the corresponding
coefficient $z(W)$ equals zero. The coefficients $z(W)$ may be
determined in various ways. A simple method to determine these
coefficients has been suggested recently by
Reinsch~\cite{reinsch:1}. His method is easily implemented in a
computer program~\cite{reinsch:1, weisstein:1}. Another method was
presented by Goldberg~\cite{goldberg:1}, and his method was
implemented in a program by Newman and Thompson~\cite{newman:1}.

The terms $z_n$ of the BCH series may be expressed as
linear combinations of nested commutators of $x$ and $y$. This was
originally shown by Baker, Campbell, and
Hausdorff~\cite{hausdorff:1}. However, explicit determination of the
coefficients was difficult for general $n$.
Dynkin~\cite{dynkin:1, dynkin:2} showed that the coefficients are in
fact easily obtained, so that the $z_n$ can be represented as
\begin{equation}\label{introzn}
  z_n~=~\frac{1}{n}\sum_{W(s_1,\ldots,s_n)} z(W)~ [[\ldots[[s_1,s_2],s_3],\ldots],s_n]
\end{equation}
where $[[\ldots[[s_1,s_2],s_3],\ldots],s_n]$ is a direct translation
of the word $W$ into a left normal nested commutator, i.e. the order
of the letters in the commutator and the word are the same. The
representation of the $z_n$ in terms of
commutators is not unique due to the Jacobi identity
$[[x,y],z]+[[y,z],x]+ [[z,x],y]=0$ and similar identities for higher
order commutators. Some of them are discussed by
Oteo~\cite{oteo:1}. Oteo also formulates a conjecture concerning
another translation of the BCH terms into a linear combination of commutators. This translation
consists of fewer terms than Dynkin's translation, but a
general proof of its  validity is not known to us.  We have checked
the validity of Oteo's conjecture up to order $n=17$ on a computer.

It was first shown by Zassenhaus~\cite{magnus:1} that there exists a
formula analogous to the BCH formula for the exponential of the sum
of two non-commuting variables $a$ and $b$,
\begin{equation}\label{zassenhaus-exp}
  \e^{a+b}~=~\e^a\cdot\e^b\cdot\prod_{n=2}^{\infty}\e^{c_n},
\end{equation}
which is known as the Zassenhaus product formula. The individual
terms $c_n$ will be called the Zassenhaus exponents in the
following. They may also be written in terms of words of length $n$
consisting of the letters $a$ and $b$, i.e.
\begin{equation}\label{introcn}
  c_n~=\sum_{W(t_1,\ldots,t_n)} c(W)~ W(t_1,\ldots,t_n)
\end{equation}
with $t_i=a$ or $t_i=b$. It is the goal of this note to provide a
simple method for the calculation of the coefficients $c(W)$ as well
as to propose a suitable computer implementation. Our method has
been developed in analogy to the procedure proposed by
Reinsch~\cite{reinsch:1} for the BCH terms.

The Zassenhaus exponents may also be obtained in terms of nested
commutators as shown  e.g. in Refs.~\cite{magnus:1, wilcox:1}.
Dynkin's theorem~\cite{dynkin:1} provides a (generally
valid) translation of words into nested commutators, if such a translation exists.
Therefore, it is possible to
directly translate the expressions for the Zassenhaus
exponents~(\ref{introcn}) into a linear combination of left normal
nested commutators as in Eq.~(\ref{introzn}). In analogy to
the conjecture by Oteo~\cite{oteo:1} for the BCH terms, we
formulate here a conjecture concerning another translation of words
into left normal nested commutators for the Zassenhaus exponents
involving fewer terms than Dynkin's translation. Using a computer, we
found this conjecture to be valid for the Zassenhaus exponents up to
order $n=17$, but at this time we cannot provide a general proof.

\section{The Zassenhaus exponents}\label{sec:zassenhaus}

In this section we state a corollary which allows a recursive
determination of the Zassenhaus exponents.

Let $\tau_1,\ldots,\tau_n$ be commuting variables. In terms of these
variables we define three upper triangular $(n+1)\times(n+1)$
matrices $H$, $K$, and $L$ with matrix elements given by
\begin{eqnarray}
  H_{ij} & = & \frac{1}{(j-i)!}\cdot\prod_{k=i}^{j-1}(1+\tau_k),~~~~~K_{ij}~\,=\,~\frac{(-1)^{i+j}}{(j-i)!}, \\
  ~ & ~ & ~~~~~~~~~L_{ij}~\,=~\,\frac{(-1)^{i+j}}{(j-i)!}\cdot\prod_{k=i}^{j-1}\tau_k
\end{eqnarray}
for $1\leq i \leq j\leq n+1$ and zero otherwise. These matrices may
be expressed as exponentials of the $(n+1)\times(n+1)$ matrices $P$
and $Q$ defined  by
\begin{equation}\label{pq}
P_{ij}\,=\,\delta_{i+1,j},~~~~~~Q_{ij}\,=\,\delta_{i+1,j}\,\tau_i
\end{equation}
where $\delta_{ij}$ is the Kronecker symbol,
\begin{equation}\label{eq:klj:inverseexp}
  H=\exp(P+Q),~~~K=\exp(-P),~~~L=\exp(-Q).
\end{equation}

Furthermore, we define an operator $U$ which operates on products
$p$ of the variables $\tau_i$
$$p~=~\tau_1^{\mu_1}\tau_2^{\mu_2}\tau_3^{\mu_3}\ldots\tau_n^{\mu_n}$$
with $\mu_i\in\{0,1\}$ for $i=1,..,n$. The operator $U$
``translates" such a product $p$ into a word consisting of letters
$a$ and $b$ according to the following rule: If $\mu_i=0$,
$\tau_i^{\mu_i}$ is replaced by an $a$, and if $\mu_i=1$,
$\tau_i^{\mu_i}$ is replaced by a $b$. The index $i$ determines  the
position of the letter in the word. E.g., for $n=6$ the product
$p~=~\tau_1^1\tau_2^0\tau_3^1\tau_4^1\tau_5^0\tau_6^0~=~\tau_1\tau_3\tau_4$
is translated as follows:
$$U(p)~=~U(\tau_1^1\tau_2^0\tau_3^1\tau_4^1\tau_5^0\tau_6^0)~=~U(\tau_1\tau_3\tau_4)~=~babbaa.$$
The operator $U$ is a vector-space isomorphism mapping the space of
polynomials in the $\tau$-variables (with $\mu_i=0$ or $\mu_i=1$)
into the space of words of length $n$.

\newtheorem{zass}{Corollary}

\begin{zass}\label{sec:zassenhaus:behauptung}
The Zassenhaus exponent $c_2$ defined in Eq. (\ref{zassenhaus-exp})
is obtained in terms of the $3\times 3$ matrices $L,~K,~H$ as
$c_2=U(L \cdot K \cdot H)_{1,3}$. For $n > 2$, the Zassenhaus
exponents $c_n$ are given in terms of the corresponding
$(n+1)\times(n+1)$ matrices as
\begin{equation}\label{eq:zassenhaus:behauptung}
  c_n ~=~ U\left(\left(\e^{-\mathcal{C}_{n-1}}\cdot\ldots\cdot\e^{-\mathcal{C}_2}\cdot L\cdot K\cdot H\right)_{1,n+1}\right)
\end{equation}
Here, $\mathcal{C}_m$ ($1<m<n$) are the Zassenhaus exponents written
in terms of the $(n+1)\times(n+1)$ matrices $P$ and $Q$, and the
index $1,n+1$ indicates the upper right element of a matrix.
\end{zass}

This corollary permits a recursive determination of the Zassenhaus
exponents. In fact, due to the special structure of the matrices $P$
and $Q$ all exponentials in Eq.~(\ref{eq:zassenhaus:behauptung}) are
obtained as finite sums, and the whole calculation can be done in a
finite amount of steps either by hand or on a computer. A suitable
computer implementation will be presented in
section~\ref{sec:zassenhaus:algorithmus}.

\subsection{Examples}

Before proving Corollary 1 we work out a number of examples.

For $n=2$ we need to use the $3\times3$ matrices given by
\begin{eqnarray*}
  L & = & \left(\begin{array}{ccc}
    1 & -\tau_1 & \frac{1}{2}\tau_1\tau_2 \\
    0 & 1 & -\tau_2 \\
    0 & 0 & 1
  \end{array}\right),~~~~K~=~\left(\begin{array}{ccc}
    1 & -1 & \frac{1}{2} \\
    0 & 1 & -1 \\
    0 & 0 & 1
  \end{array}\right), \\
  H & = & \left(\begin{array}{ccc}
    1 & (1+\tau_1) & \frac{1}{2}(1+\tau_1)(1+\tau_2) \\
    0 & 1 & (1+\tau_2) \\
    0 & 0 & 1
  \end{array}\right).
\end{eqnarray*}
Then,
$$L\cdot K\cdot H~=~\left(\begin{array}{ccc}
  1 & 0 & \frac{1}{2}\tau_1-\frac{1}{2}\tau_2 \\
  0 & 1 & 0 \\
  0 & 0 & 1
\end{array}\right),$$
and the second Zassenhaus exponent takes the form
\begin{eqnarray}\label{c2}
  c_2 & = & U(L\cdot K\cdot H)_{1,n+1} \,~=~\, U\left(\frac{1}{2}\tau_1-\frac{1}{2}\tau_2\right) \nonumber\\
  ~ & = & \frac{1}{2}\,U(\tau_1^1\tau_2^0-\tau_1^0\tau_2^1) \,~=~\, \frac{1}{2}(ba-ab).
\end{eqnarray}

For $n=3$ we need to use the $4\times4$ matrices
\begin{eqnarray*}
  L & = & \left(\begin{array}{cccc}
    1 & -\tau_1 & \frac{1}{2}\tau_1\tau_2 & -\frac{1}{6}\tau_1\tau_2\tau_3\\
    0 & 1 & -\tau_2 & \frac{1}{2}\tau_2\tau_3\\
    0 & 0 & 1 & -\tau_3 \\
    0 & 0 & 0 & 1
  \end{array}\right), \\
  K & = & \left(\begin{array}{cccc}
    1 & -1 & \frac{1}{2} & -\frac{1}{6}\\
    0 & 1 & -1 & \frac{1}{2}\\
    0 & 0 & 1 & -1 \\
    0 & 0 & 0 & 1
  \end{array}\right), \\[4pt]
  H & = & \left(\begin{array}{cccc}
    1 & (1+\tau_1) & \frac{1}{2}(1+\tau_1)(1+\tau_2) & \frac{1}{6}\sum\limits_{i=1}^3(1+\tau_i) \\
    0 & 1 & (1+\tau_2) & \frac{1}{2}\sum\limits_{i=2}^3(1+\tau_i)\\
    0 & 0 & 1 & (1+\tau_3) \\
    0 & 0 & 0 & 1
  \end{array}\right)
\end{eqnarray*}
and the matrices $P$ and $Q$ defined in Eq.~(\ref{pq})
$$P\,=\left(\begin{array}{cccc}
  0 & 1 & 0 & 0 \\
  0 & 0 & 1 & 0 \\
  0 & 0 & 0 & 1 \\
  0 & 0 & 0 & 0
\end{array}\right)~~~~~~\textnormal{and}~~~~~~Q\,=\left(\begin{array}{cccc}
  0 & \tau_1 & 0 & 0 \\
  0 & 0 & \tau_2 & 0 \\
  0 & 0 & 0 & \tau_3 \\
  0 & 0 & 0 & 0
\end{array}\right).$$
It follows from Eq. (\ref{c2}) that
\begin{eqnarray*}
  \e^{-C_2} & = & \exp\left(-\frac{1}{2}(Q\cdot P-P\cdot Q)\right) \\
  ~ & = & \left(\begin{array}{cccc}
    1 & 0 & \frac{1}{2}(\tau_2-\tau_1) & 0 \\
    0 & 1 & 0 & \frac{1}{2}(\tau_3-\tau_2) \\
    0 & 0 & 1 & 0 \\
    0 & 0 & 0 & 1
  \end{array}\right).
\end{eqnarray*}
Therefore,  the third Zassenhaus exponent takes the form
\begin{eqnarray*}
  c_3 & = & U\left(\e^{-C_2}\cdot L\cdot K\cdot H\right)_{1,n+1} \\
  ~ & = & U\left(\frac{2}{3}\tau_1\tau_3-\frac{1}{3}\tau_2-\frac{1}{3}\tau_2\tau_3-\frac{1}{3}\tau_1\tau_2+\frac{1}{6}\tau_1+\frac{1}{6}\tau_3\right) \\
  ~ & = & \frac{2}{3}bab-\frac{1}{3}aba-\frac{1}{3}abb-\frac{1}{3}bba+\frac{1}{6}baa+\frac{1}{6}aab.
\end{eqnarray*}

In an analogous way one obtains for  $c_4$
\begin{eqnarray*}
  c_4 & = & -\frac{1}{24}aaab+\frac{1}{8}aaba+\frac{1}{8}aabb-\frac{1}{8}abaa \\[4pt]
  ~ & ~ & -\frac{1}{4}abab-\frac{1}{8}abbb+\frac{1}{24}baaa+\frac{1}{4}baba \\[4pt]
  ~ & ~ & +\frac{3}{8}babb-\frac{1}{8}bbaa-\frac{3}{8}bbab+\frac{1}{8}bbba.
\end{eqnarray*}

These results all agree with the standard results given in the
literature, see e.g.~Ref.~\cite{quesne:1}.

\section{Proof of Corollary 1}\label{sec:zassenhaus:beweis}

The $(n+1)\times(n+1)$ matrices $P$ and $Q$ have non-zero elements
only in their first  superdiagonal. A product of  $m$ factors $P$ or
$Q$ is a $(n+1)\times(n+1)$ matrix which contains non-zero elements
only in the $m\,$th superdiagonal. In particular,  a product of
 $n$ factors $P$ or $Q$ has only one non-zero element in
its upper right corner. A product of $k$ factors $P$ or $Q$ with
$k>n$ is a null matrix.

Each Zassenhaus exponent $c_n$ is a linear combination of words of
length $n$. As a consequence, $\mathcal{C}_n$  is a
$(n+1)\times(n+1)$ matrix, which has a non-zero entry only in its
upper right corner, and for each  $n\in\NN$
\begin{equation}\label{eq:zassenhaus:produkt}
  \e^{P+Q}~=~\e^P\cdot\e^Q\cdot\prod_{i=1}^{n} \e^{\mathcal{C}_i}
\end{equation}
is obtained in terms of a finite product with $n$ factors.
Therefore, one obtains
\begin{equation}
  \e^{\mathcal{C}_n} ~=~ \e^{-\mathcal{C}_{n-1}}\cdot\ldots\cdot\e^{-\mathcal{C}_{2}}\cdot\e^{-Q}\cdot\e^{-P}\cdot\e^{P+Q}.
\end{equation}
The exponentials can be calculated from a finite sum.

Since  $\mathcal{C}_n$ only has one non-zero element in its upper
right corner, it holds  that $\e^{\mathcal{C}_n}=I+\mathcal{C}_n$,
and we find
\begin{eqnarray}\label{eq:zassenhaus:glung}
\left(\mathcal{C}_n\right)_{1,n+1} &=&\left(\sum c(A_1 \ldots A_n)~
A_1\ldots A_n\right)_{1,n+1}\nonumber\\
  &=& \left(\e^{\mathcal{C}_n}\right)_{1,n+1}  \nonumber\\
 &=&
  \left(\e^{-\mathcal{C}_{n-1}}\cdot\ldots\cdot\e^{-\mathcal{C}_{2}}\cdot\e^{-Q}\cdot\e^{-P}\cdot\e^{P+Q}\right)_{1,n+1},
\end{eqnarray}
where $A_i=P$ or $A_i=Q$, and  the sum runs over all  different
matrix products $A_1\ldots A_n$. The next step is to show that the
upper-right element of a product matrix $A_1\ldots A_n$ is given by
a product of $\tau_i$, and that the indices on the $\tau_i$
variables determine the positions of the $Q$'s in the matrix product
$A_1\ldots A_n$. This has been shown in Ref.~\cite{reinsch:1} and
 is not repeated here. Applying the operator $U$ on
this product of $\tau_i$ then transforms the result
Eq.~(\ref{eq:zassenhaus:glung}) into a linear combination of words
in terms of the letters $a$ and $b$. This proves Corollary 1.

\section{Computer Implementation}\label{sec:zassenhaus:algorithmus}

The following \textit{Mathematica} program implements Corollary 1.
Calling it will return the Zassenhaus exponent $c_n$ in terms of the
variables $a$ and $b$.

The program consists of three parts: First the matrices $L,~K,~H,~P$
and $Q$ are defined. Then the product of exponentials as required by
Corollary 1 is calculated (starting from $n=2$), and finally the
translation $U$ is implemented. The program works with strings in
order to prevent \textit{Mathematica} from sorting the words
alphabetically.

\begin{small}
\begin{verbatim}
ZH[n_,a_,b_] := Module[{C,L,K,H,P,Q,m,t,r,i,j,k,u,z},
C[2]=(t[1]^2 t[2])/2-(t[1] t[2]^2)/2;

For[m = 2, m <= n, m++,
L = Table[(-1)^(i+j)/(j-i)! Product[t[k],{k,i,j-1}],{i,m+1},{j,m+1}];
K = Table[(-1)^(i+j)/(j-i)!, {i,m+1},{j,m+1}];
H = Table[1/(j-i)! Product[(1+t[k]),{k,i,j-1}],{i,m+1},{j,m+1}];
P = Table[KroneckerDelta[i+1,j],{i,m+1},{j,m+1}];
Q = Table[KroneckerDelta[i+1,j] t[i],{i,m+1},{j,m+1}];

C[m] = Expand[
  ((Dot @@ Table[MatrixExp[-Sum[
  r = (List @@ C[m-u][[z]]) /. {t[i_] -> P, t[i_]^2 -> Q};
  r[[1]] (Dot @@ Take[r,-Length[r]+1]),
  {z,Length[C[m-u]]}]],
  {u,1,m-2}]).L.K.H)[[1, m+1]] Product[t[j],{j,m}]]];

Sum[r = (List @@ C[n][[k]])
         /. {t[i_] -> ToString[a], t[i_]^2 -> ToString[b]};
    r[[1]] (StringJoin @@ Take[r,-n]), {k, Length[C[n]]} ]  ];
\end{verbatim}
\end{small}

More elegant but less readable \textit{Mathematica}
implementations than the one given above are possible, e.g. using \verb"NestList" instead of
a \verb"For" loop.
Since the Zassenhaus exponents for larger $n$ are rather lengthy
expressions, the computer needs a significant amount of memory for
this calculation. On a standard personal computer with 2 GB of
memory we could obtain the Zassenhaus exponents up to $n=17$ within
about one hour of computer time.

\section{Expression in terms of commutators}

In the introduction we briefly discussed Dynkin's
translation~\cite{dynkin:1, dynkin:2} of words into commutators, which is applicable,
whenever such a translation exists. Since it is known that a representation in terms of commutators exists
for the Zassenhaus exponents~\cite{magnus:1, wilcox:1}, we may directly use
Dynkin's prescription in order to obtain an explicit representation of the Zassenhaus exponents
in terms of commutators. We checked the validity of this procedure using a computer up to order $n=17$.
A translation for the BCH terms into commutators involving fewer terms
than Dynkin's prescription was proposed by Oteo~\cite{oteo:1}.
To our knowledge the validity of this translation has never been proved in general.
Oteo showed it to be valid up to order $n=10$, and we checked this conjecture up to order $n=17$ on
a computer.

In analogy to Oteo's prescription for the BCH terms, we now write down an expression for the Zassenhaus
exponents
\begin{equation}
c_n~=\sum_{W(t_1,..,t_n)}c(W)~ W(t_1,..,t_n)
\end{equation}
in terms of commutators. The words W consist of letters a and b;
$n_a(W)$ counts the number of $a$'s in that word. Analogously,
$n_b(W)=n-n_a(W)$ counts the number of $b$'s in the word $W$. We
conjecture that the Zassenhaus exponent $c_n$ may be expressed in
terms of the left normal commutator as follows
\begin{equation}
  c_n~=~\sum_{W(t_1,..,t_n)\atop t_1=b,\,t_2=a}\frac{c(W)}{n_b(W)}~
  [[..[[t_1,t_2],t_3],...],t_n].
\end{equation}
Here we only sum over words starting with the letters $ba$.
Similarly, one may write
\begin{equation}
  c_n~=~\sum_{W(t_1,..,t_n)\atop t_1=a,\,t_2=b}\frac{c(W)}{n_a(W)}~ [[..[[t_1,t_2],t_3],...],t_n]
\end{equation}
and only sum over words starting with the letters  $ab$.

We checked this conjecture up to order $n=17$ using
\textit{Mathematica} and compared results up to order $n=6$ with
expressions given in the literature, e.g. in Ref.~\cite{quesne:1}.

\subsection{Example}

For $n=4$ one finds the following representation of the Zassenhaus
exponent in terms of words
\begin{eqnarray*}
  c_4 & = & -\frac{1}{24}aaab+\frac{1}{8}aaba+\frac{1}{8}aabb-\frac{1}{8}abaa \\[4pt]
  ~ & ~ & -\frac{1}{4}abab-\frac{1}{8}abbb+\frac{1}{24}baaa+\frac{1}{4}baba \\[4pt]
  ~ & ~ & +\frac{3}{8}babb-\frac{1}{8}bbaa-\frac{3}{8}bbab+\frac{1}{8}bbba.
\end{eqnarray*}
According to our conjecture this may be translated into nested
commutators as
\begin{eqnarray*}
  c_4 & = & \frac{1}{24}[[[b,a],a],a]+\frac{1}{8}[[[b,a],b],a]+\frac{1}{8}[[[b,a],b],b] \\
  ~ & = & -\frac{1}{24}[[[a,b],a],a]-\frac{1}{8}[[[a,b],a],b]-\frac{1}{8}[[[a,b],b],b].
\end{eqnarray*}
This result agrees with results given in the literature (e.g. in
Ref. [10]).

\section{Conclusion}

We developed a method for the calculation of the Zassenhaus
exponents $c_n$ in the Zassenhaus product formula
$\e^{a+b}=\e^a\cdot\e^b\cdot\prod_{n=2}^{\infty}\e^{c_n}$ for
non-commuting $a$ and $b$.
The method is given in Corollary 1, from which we
obtain the Zassenhaus exponents in terms of words. It appears that
our method is simpler and faster than previous methods (see e.g.
Ref.~\cite{wilcox:1}). We provide a suitable computer implementation.

We would like to mention that the method presented here can be
easily generalized to the Zassenhaus product formula for
$q$-deformed exponentials. (The Zassenhaus formula for $q$-deformed
exponentials is discussed e.g. in Ref.~\cite{quesne:1} and
references therein.)

Furthermore, we formulated a conjecture on how to translate the Zassenhaus exponents given in
terms of words into a form in terms of left normal nested
commutators. This representation involves fewer terms than a translation based on Dynkin's
theorem~\cite{dynkin:2} and has been found to be valid for Zassenhaus
exponents up to order $n=17$ using a computer. We expect a proof of our
conjecture to be possible along the lines of Dynkin's proof for his representation
of the BCH formula in terms of commutators. This proof shows essentially by direct calculation
that the conjectured commutator representation is equivalent to the representation in terms of words.
We will address this issue in a forthcoming paper.

\acknowledgments We acknowledge a useful correspondence with M.
Reinsch and helpful conversations with M. Reginatto. D. S.
acknowledges an internship at the Physikalisch-Technische
Bundesanstalt.

%\bibliographystyle{unsrt}
%\bibliography{../references}

\small
\begin{thebibliography}{10}

\bibitem{reinsch:1}
M.W. Reinsch,
\newblock {\em J. Math. Phys.}, {\bf 41}, 2434 (2000).

\bibitem{weisstein:1}
E.~W. Weisstein,
\newblock {\em ``Baker-Campbell-Hausdorff Series." From Mathworld -- A Wolfram
  Web Resource.}\\ http://mathworld.wolfram.com/Baker-Campbell-HausdorffSeries.html.

\bibitem{goldberg:1}
K.~Goldberg,
\newblock {\em Duke Math. J.}, {\bf 23}, 13 (1956).

\bibitem{newman:1}
M.~Newman and R.~C. Thompson,
\newblock {\em Math. Comp.}, {\bf 48}, 265 (1987).

\bibitem{hausdorff:1}
H.F. Baker, Proc. London Math. Soc., {\bf 3}, 24 (1904),
J.E. Campbell, Proc. London Math. Soc., {\bf 29}, 14 (1898),
F.~Hausdorff,
\newblock {\em Ber. Verh. S{\"a}chs. Akad. Wiss., Leipzig, Math.-Naturw. Kl.},
  {\bf 58}, 19 (1906).

\bibitem{dynkin:1}
E.B. Dynkin,
\newblock {\em Doklady Akademii Nauk SSSR}, {\bf 57}, 323 (1947),
\newblock [in Russian]. \newblock {An English translation may be found in ``Selected papers
  of E.B. Dynkin with commentary", E.B. Dynkin, A.A. Yushkevich, G. M. Seitz, A.
  L. Onishchik, editors. American Mathematical Society, Providence, R.I., and
  International Press, Cambridge, Mass., 2000.}


\bibitem{dynkin:2}
E.B. Dynkin,
\newblock {\em Mathematicheskii Sbornik}, {\bf 25(67)}, 155 (1949),
\newblock [in Russian].

\bibitem{oteo:1}
J.A. Oteo,
\newblock {\em J. Math. Phys.}, {\bf 32}, 419 (1991).

\bibitem{magnus:1}
W.~Magnus,
\newblock {\em Comm. Pure Appl. Math.}, {\bf 7}, 649
  (1954).

\bibitem{wilcox:1}
R.M. Wilcox,
\newblock {\em J. Math. Phys.}, {\bf 8}, 962 (1967).

\bibitem{quesne:1}
C.~Quesne,
\newblock {\em Intern. J. Theo. Phys.}, {\bf 43}, 545
  (2004).

\end{thebibliography}

\end{document}